\documentclass[lettersize,journal]{IEEEtran}
\usepackage{amsmath,amsfonts}
\usepackage{algorithmic}
\usepackage{algorithm}
\usepackage{array}
\usepackage[caption=false,font=normalsize,labelfont=sf,textfont=sf]{subfig}
\usepackage{textcomp}
\usepackage{stfloats}
\usepackage{url}
\usepackage{cite}
\usepackage{verbatim}
\usepackage{graphicx}
\usepackage{setspace}
\usepackage{booktabs}
\usepackage{enumerate}
\usepackage{diagbox}
\usepackage{arydshln}
\usepackage{accents}
\usepackage{color}
\usepackage{balance}
\usepackage{bm}

\def\BibTeX{{\rm B\kern-.05em{\sc i\kern-.025em b}\kern-.08em
    T\kern-.1667em\lower.7ex\hbox{E}\kern-.125emX}}
\usepackage{balance}
\begin{document}
\title{{Block-Wise Index Modulation and Receiver Design for High-Mobility OTFS Communications}}
\author{Mi Qian, \textit{Student~Member,~IEEE}, Fei Ji, \textit{Member,~IEEE}, \\Yao Ge, \textit{Member,~IEEE}, Miaowen Wen, \textit{Senior~Member,~IEEE}, \\Xiang Cheng, \textit{Fellow,~IEEE}, and H. Vincent Poor, \textit{Life~Fellow,~IEEE} 
	
\thanks{This work was supported in part by the National Natural Science Foundation of China (NSFC) under Grant 62192712, in part by the Guangdong Basic and Applied Basic Research Foundation under Grants 2023A1515030118 and 2021B1515120067, and in part by the U.S National Science Foundation under Grant CNS-2128448 \textit{(Corresponding authors: Miaowen Wen, Yao Ge.)}}

\thanks{Mi Qian, Fei Ji, and Miaowen Wen are with the School of Electronic and Information Engineering, South China University of Technology, Guangzhou 510641, China (e-mail: eemqian@mail.scut.edu.cn, \{eefeiji, eemwwen\}@scut.edu.cn).}
\thanks{Yao Ge is with the Continental-NTU Corporate Lab, Nanyang Technological University, Singapore (e-mail: yao.ge@ntu.edu.sg).}
\thanks{Xiang Cheng is with the State Key Laboratory of Advanced Optical Communication Systems and Networks, Department of Electronics, School of Electronics Engineering and Computer Science, Peking University, Beijing 100871, China (e-mail: xiangcheng@pku.edu.cn).}
\thanks{H. Vincent Poor is with the Department of Electrical and Computer Engineering, Princeton University, Princeton, NJ 08544, USA (e-mail: poor@princeton.edu).}
}


\maketitle
\vspace{-10mm}
\begin{abstract}
As a promising technique for high-mobility wireless communications,  
orthogonal time frequency space (OTFS) has been proved to enjoy excellent advantages with respect to traditional orthogonal frequency division multiplexing (OFDM). Although multiple studies have considered index modulation (IM) based OTFS (IM-OTFS) schemes to further improve system performance, a challenging and open problem is the development of effective IM schemes and efficient
receivers for practical OTFS systems that must operate in the presence of channel delays and Doppler shifts. In this paper, we propose two novel block-wise IM schemes for OTFS systems, named delay-IM with OTFS (DeIM-OTFS) and Doppler-IM with OTFS (DoIM-OTFS), where a block of delay/Doppler resource bins are activated simultaneously. Based on a maximum likelihood (ML) detector, we analyze upper bounds on the average bit error rates for the proposed DeIM-OTFS and DoIM-OTFS schemes, and verify their performance advantages over the existing IM-OTFS systems. We also develop a multi-layer joint symbol and activation pattern detection (MLJSAPD) algorithm and a customized message passing detection (CMPD) algorithm for our proposed DeIM-OTFS and DoIM-OTFS systems with low complexity.  Simulation results demonstrate that our proposed MLJSAPD and CMPD algorithms can achieve desired performance with robustness to the imperfect channel state information (CSI).
\end{abstract}


\section{Introduction}
Nowadays, a large number of wireless applications such as communication with high-speed trains and unmanned autonomous vehicles are emerging. Accordingly, it is important to have high data rate and low latency communications to satisfy the fast-growing requirements expected in the future. Orthogonal frequency division multiplexing (OFDM) modulation is prevalent in today's wireless systems as it is able to provide high spectral efficiency and is easy to implement \cite{7936676,8049521,6587554}. However, for time-varying channels with large Doppler spread, OFDM can suffer significant performance degradation due to the loss of orthogonality or inter-carrier-interference~(ICI). 
                                                                                            
To cope with high-mobility scenarios, a new modulation scheme referred to as orthogonal time frequency space (OTFS) has been proposed \cite{7925924,9508932,Yuan2022ASO}, which can achieve significant performance improvement over OFDM modulation. OTFS can exploit the diversity gain from both the delay and Doppler dimensions of a mobile wireless channel since all transmitted symbols can be multiplexed in the delay-Doppler domain and spread over the time-frequency domain \cite{8424569,8686339,9404861,8599041,9082873,8757044}. Furthermore, OTFS can convert the time-varying channel into a two-dimensional (2D) quasi-time-invariant channel in the delay-Doppler domain, which significantly reduces the complexity of channel estimation \cite{8727425,8671740,9864300} and symbol detection \cite{9492800,8516353,8701706,8892482,9321356,8859227,8918014} at the receiver. Attracted by its advantages, a number of studies of OTFS have examined it in concert with non-orthogonal multiple access (NOMA) \cite{9354639}, millimeter wave (mmWave) communication systems \cite{8746382}, and integrated sensing and communication \cite{9557830}. In \cite{9354639}, the authors investigated an OTFS-based NOMA configuration in which each group of co-channel mobile and stationary users is modulated by OTFS. The work in \cite{8746382} addressed the effect of oscillator phase noise on the performance of mmWave OTFS systems, where oscillator phase noise and Doppler shifts are typically high. The authors of \cite{9557830} proposed a novel integrated sensing and communication-assisted OTFS transmission scheme in vehicle-to-infrastructure scenarios, which reduces the hardware cost as well as the demand on spectral resources.

Index modulation (IM), whicn enjoys high spectral and energy efficiency, is a promising modulation technique for next generation wireless networks \cite{7547943,9380189}. In IM schemes, information bits are transmitted not only by $M$-ary signal constellations but also by the indices of transmission entities. Many kinds of transmission entities, such as antennas \cite{5165332}, OFDM subcarriers \cite{6841601,7583706} and frequency slots \cite{7676245}, can be used for carrying index bits without extra energy consumption. 

Recognizing the superiority of IM, index modulation based orthogonal time frequency space (IM-OTFS) \cite{9129380} has been recently proposed to improve the bit error rate (BER) performance for high-mobility communication scenarios. Specifically, the index bits are transmitted by the indices of the activated OTFS delay-Doppler resources, where the active resource bins are independently randomly selected. To further improve the system performance, OTFS with dual-mode index modulation (OTFS-DM-IM) was proposed in \cite{9335633}, which provides a desired trade-off between transmission reliability and spectral efficiency (SE). To effectively decode the index bits and constellation bits, several detectors have also been proposed in the literature. In \cite{9129380}, a minimum mean squared error with maximum likelihood (MMSE-ML) detector was proposed, where the MMSE criterion was employed for the detection of constellation bits and index bits, and the ML principle is utilized to detect the indices information. In \cite{9335633}, a modified log likelihood ratio (LLR) detector based on the minimum Hamming distance was investigated to improve the BER performance. However, the performance analysis for the designed schemes and detectors in \cite{9129380} and \cite{9335633} only considers ideal bi-orthogonal OTFS pulses and requires mobile channels exhibiting on-the-grid delays and Doppler shifts, which are unrealistic assumptions in practical OTFS system deployment. 

On the other hand, the channel delay and Doppler shifts will cause severe inter-symbol interference (ISI) in high mobility OTFS communications. The existing IM-OTFS systems \cite{9129380,9335633} only activate independent delay-Doppler resources and cannot determine the active and inactive resources accurately at the receiver, leading to an inevitable performance loss. Therefore, it is necessary to develop more efficient and reliable IM schemes for OTFS transmissions by considering the effects caused by the channel delays and Doppler spreads. To date, there has been no relevant work taking these factors into account.   

In this paper, we propose effective block-wise IM schemes and develop efficient receiver algorithms for OTFS systems to alleviate the delay-Doppler channel effects. We also dispense with the impractical assumption that the channel delays and Doppler shifts are on the OTFS sampling grid, and analyze the performance of our proposed schemes. Our contributions in this paper are summarized as follows:
\begin{itemize}
	\item{We propose two effective block-wise IM schemes for OTFS systems, denoted as delay-IM with OTFS (DeIM-OTFS) and Doppler-IM with OTFS (DoIM-OTFS), where a block of delay/Doppler resource bins are activated simultaneously. The proposed schemes can operate with practical rectangular pulses and work well for the practical scenarios where the channel delay and Doppler shifts do not necessarily land on the OTFS delay-Doppler sampling grid.}
	\item{We derive asymptotically tight BER upper bounds for the DeIM-OTFS and DoIM-OTFS schemes with the optimal ML detectors. The performance improvement of our proposed block-wise IM schemes for OTFS is also verified in contrast to the existing IM-OTFS schemes.}
	\item{We develop a multi-layer symbol and activation pattern detection (MLJSAPD) algorithm and a customized message passing detection (CMPD) algorithm for the proposed DeIM-OTFS and DoIM-OTFS schemes. The MLJSAPD introduces a new layer in the factor graph to further track the activated blocks of the transmitted symbols. The CMPD algorithm can effectively identify the active resource units by considering the active probability of each resource unit during the iterations. }
	\item{Simulation results demonstrate that the proposed MLJSAPD and CMPD algorithms can achieve desired performance with relatively low complexity for both DeIM-OTFS and DoIM-OTFS systems, and also robustness against imperfect channel state information (CSI). }
\end{itemize}

The rest of this paper is organized as follows. In Section II, we first introduce our proposed block-wise IM schemes and also describe the corresponding system model.
In Section III, we analyze the theoretical BER upper bounds of the proposed DeIM-OTFS and DoIM-OTFS schemes with the ML detector. The proposed low-complexity MLJSAPD and CMPD detectors are described in Section IV. Simulation results are presented in Section V. Finally, Section VI concludes the paper.

$Notation:$ $\left ( \cdot  \right )$$^{\rm{T}}$, $\left ( \cdot  \right )$$^{*}$, $\left ( \cdot  \right )$$^{\rm{H}}$, and $\left \| \cdot  \right \|$ denote the transpose, conjugate, Hermitian operations, and Euclidean norm of a matrix, respectively. $\left \lfloor . \right \rfloor$ denotes the integer floor operator. $[ \cdot ]_{m}$ denotes the mod-$m$ operation. $\mathbb{C}$ and $\mathbb{Z}$ denote the set of complex numbers and positive integers, respectively. $S$ is the constellation set. $\rm{C}$$(n ,k )$ denotes the binomial coefficient that chooses $k$ out of $n$. $\mathbb{E}(\cdot)$, det$\left ( \cdot  \right )$, diag$(\cdot )$, and $Q(.)$ denote the expectation, determinant, diagonal matrix, and Gaussian $Q$-function, respectively.  

\section{System Model}

In this section, we briefly introduce our proposed block-wise IM schemes for OTFS and also present the corresponding system model, which are shown in Fig.~1 and Fig.~2, respectively.

A 2D lattice in the time-frequency plane is sampled at interval $T$ (seconds) and $\Delta f=1/T$ (Hz), respectively, i.e., $\Lambda =\left \{ \left ( m\Delta f , nT\right ),m=0,\ldots, M-1;n=0,\ldots,N-1\right \}$ for $M\in \mathbb{Z}, N\in \mathbb{Z}$. Here, $M$ and $N$ represent the total available numbers of subcarriers and time slots, respectively. $\Delta f$ and $T$ are chosen larger than the maximum Doppler frequency shift $\nu _{max}$ and maximal channel delay spread $\tau _{max}$, respectively. Thus, the corresponding delay-Doppler plane is described as an information grid, i.e., $\Gamma =\left \{ \left ( \frac{\ell}{M\Delta f},\frac{k}{NT} \right ),\ell=0,\ldots,M-1 ;k=0,\ldots,N-1\right \}$, where the sampling time $1/M\Delta f$ and sampling frequency $1/NT$ are referred to as the delay resolution and the Doppler resolution of the delay-Doppler grid, respectively. 

\subsection{Proposed Block-wise IM Schemes for OTFS}
Unlike the conventional random IM schemes applied in OTFS systems \cite{9129380,9335633}, our proposed DeIM-OTFS and DoIM-OTFS schemes activate a block of delay/Doppler resource bins simultaneously, which can help to further improve the receiver performance and combat the effect of high mobility time-varying channels.

Let us consider a total number of $\mathcal{B}$ information bits for transmission in each OTFS frame. The OTFS frame is split into $J$ subframes, each of which is composed of an $\widehat{M} \times \widehat{N}$ signal matrix. $\widehat{M}$ and $\widehat{N}$ denote the numbers of resource units in the delay dimension and Doppler dimension for each subframe, respectively. Let ${\widehat{\ell}}=0,\ldots,\widehat{M}-1$ and $\widehat{k}=0,\ldots,\widehat{N}-1$ represent indexes of delay and Doppler resource bins for each subframe, respectively. The total number of subframes is given by $J=\overline{M}\hspace{0.2mm} \overline{N}$, where $\overline{M}=M/\widehat{M}$ and $\overline{N}=N/\widehat{N}$, respectively. We denote the $\beta$-th subframe as $G[\beta]$, where $\beta={\overline{\ell}}+\overline{M}\hspace{0.2mm}\overline{k}+1$ with ${\overline{\ell}}=0,\ldots,\overline{M}-1$ and $\overline{k}=0,\ldots,\overline{N}-1$.
As shown in Fig.~1, each OTFS frame consists of $\{G[1],G[2],\ldots,G[\beta],\ldots,G[J]\}$ subframes. For each subframe, our proposed block-wise index modulator processes $p=\mathcal{B}/J$ bits in the delay-Doppler domain. These $p$ information bits are then divided into two parts: the first $p_1$ bits are transferred to the index selector to decide the active resource units; the remaining $p_2$ bits are mapped to the constellation symbols and placed on active resource units. The details of the proposed DeIM-OTFS and DoIM-OTFS schemes are respectively described as follows:

\begin{figure}
	\centering
	\includegraphics[width=3.6in,height=2.9in]{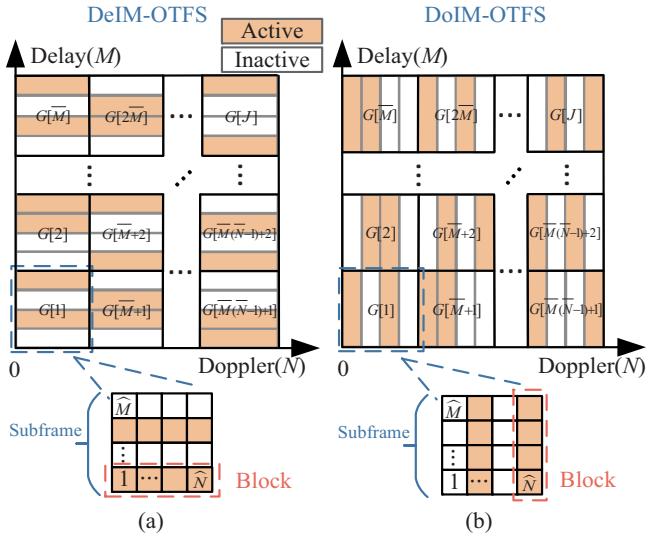}
	\caption{A snapshot of the delay-Doppler resource bins for the proposed DeIM-OTFS and DoIM-OTFS schemes.}
	\label{fig:fig1}
\end{figure}

\begin{enumerate}{}{}
	\item{DeIM-OTFS: For the DeIM-OTFS scheme, each subframe is divided into $\widehat{M}$ blocks along the delay dimension with $\widehat{N}$ Doppler resource units in each block, as shown in Fig. 1(a). We activate the resource units based on blocks according to the index bits, i.e., a block of delay resource bins are activated simultaneously. We assume the number of active blocks in each transmitted subframe is $\widehat{k}$, such that there are C($\widehat{M},\widehat{k}$) possible index combinations of active indices and $\widehat{k}\widehat{N}$ active resource units in each subframe for given $\widehat{M}$, $\widehat{N}$ and $\widehat{k}$. Therefore, the total numbers of index bits and constellation bits for each OTFS frame are given by $m_1=p_1 J= \lfloor \log_{2}  ( {\rm C}({\widehat{M}},\widehat{k}) ) \rfloor J$ and $m_2=p_2 J= \widehat{k}\log_{2}  (M_c) \widehat{N}J$, respectively, where $M_c$ represents the modulation order. The SE of the DeIM-OTFS scheme can be calculated as $S_E=(\log_{2}  ( {\rm C}(\widehat{M},\widehat{k})  )+\widehat{k}\log_{2}({M_c})\widehat{N})/(\widehat{M}\widehat{N})$. For example, in each subframe, the resource units of the first and second blocks are active if the indices of $ \{ 1,2  \}$ are selected, while the remaining inactive resource units are set to zero. }
	\item{DoIM-OTFS: For the DoIM-OTFS scheme, each subframe is divided into $\widehat{N}$ blocks along the Doppler dimension with $\widehat{M}$ delay resource units in each block, as shown in Fig. 1(b). We activate a block of Doppler resource bins simultaneously according to the index bits. For given $\widehat{M}$, $\widehat{N}$ and $\widehat{k}$, there are totally C($\widehat{N},\widehat{k}$) possible index combinations of active indices and $\widehat{k}\widehat{M}$ active resource units in each subframe. The total numbers of index bits and constellation bits in each OTFS frame are given by $m_1=\lfloor \log_{2}  ( {\rm C}({\widehat{N}},\widehat{k}))\rfloor J$ and $m_2= \widehat{k}(\log_{2}  M_c)\widehat{M} J$, respectively. The SE of the DoIM-OTFS scheme can be calculated similar to the DeIM-OTFS scheme, given by $S_E=(\log_{2}  ( {\rm C}(\widehat{N},\widehat{k}) )+\widehat{k}\log_{2}({M_c})\widehat{M})/(\widehat{M}\widehat{N})$.}
\end{enumerate}

We assume that the signal constellation symbols are normalized to have unit average power. A look-up table example is presented in Table I with parameters $\widehat{M}=4$, $\widehat{N}=4$ and $\widehat{k}=2$. Since C$(4,2)=6$, we select four index combinations out of six by abandoning the other two~cases.

\renewcommand\arraystretch{1.2}
\begin{table}[!t]
	\caption{A Look-Up Table Example for $\widehat{M}=\widehat{N}=4$, and $\widehat{k}=2$. \label{tab:table1}}
	\centering
	\begin{tabular}{|p{2.3cm}<{\centering}|p{2.3cm}<{\centering}|}
		\hline
		\text{Index bits} & \text{Indices}\\
		\hline
		\hline
		[0 0] & $\left \{ 1, 2 \right \}$\\
		\hline
		[0 1] & $\left \{ 1, 3 \right \}$\\
		\hline
		[1 0] & $\left \{ 1, 4 \right \}$\\
		\hline
		[1 1] & $\left \{ 2, 3 \right \}$\\
		\hline
	\end{tabular}
\end{table}

\subsection{Transmitter Model}
The transmitter and receiver structures of our proposed DeIM-OTFS/DoIM-OTFS system are depicted in Fig.~2. At the transmitter, the modulated signal in the $\ell$-th delay and $k$-th Doppler grid for $\ell=0,\ldots,M-1$ and $k=0,\ldots,N-1$ is given by $X[\ell,k]\in\{0,S\}$. According to the proposed DeIM-OTFS/DoIM-OTFS scheme, the delay-Doppler signal $\mathbf {X}\in \mathbb{C}^{M\times N}$ can be generated. Then, the corresponding delay-Doppler symbols $\textbf {X}$ are converted into the time-frequency domain by using the 2D inverse symplectic finite Fourier transform (ISFFT),

\begin{equation}
\overline{\mathbf{X}}=\mathbf{F}_{M} \mathbf{X} \mathbf{F}_{N}^{\text{H}},
\end{equation}
where $\mathbf{F}_{M} $ and $\mathbf{F}_{N} $ denote the normalized discrete Fourier transform (DFT) matrices of size $M \times M$ and size $N \times N$, respectively. The time-frequency domain samples \{$\overline{X}[m,n],m = 0,\ldots,M-1 ;n = 0,\ldots,N-1$\} are transmitted over an OTFS frame with duration $T_f=NT$ and occupies a bandwidth of $B=M \Delta f$. After ISFFT, the time-frequency signal $\overline{\mathbf{X}}$ is modulated through the Heisenberg transform by utilizing a transmit rectangular pulse $g_{tx}(t)$. Thus, the resulted time domain sampled signal $\mathbf{s} \in \mathbb{C}^{M N \times 1}$ can be written as
\begin{align}
s[u]=\sum_{n=0}^{N-1} \sum_{m=0}^{M-1} & \overline{X}[m, n] g_{t x}\left(u T_{s}-n T\right) e^{j 2 \pi m \Delta f\left(u T_{s}-n T\right)}, \nonumber\\
&u=0, \ldots, M N-1,
\end{align}
where $T_{s}=1 / M \Delta f$ denotes the symbol spaced sampling interval.

\subsection{Channel Model}
To eliminate the inter-frame interference, a cyclic prefix (CP) of length no shorter than the maximal channel delay spread is appended to the front of the time domain signal $\textbf{s}$. Then, $\textbf{s}$ enters the multipath fading channels after passing through a transmit filter, the channel impulse response $h[u, p]$ is characterized as
\begin{figure*}
	\centering
	\includegraphics[width=5.2in,height=3.3in]{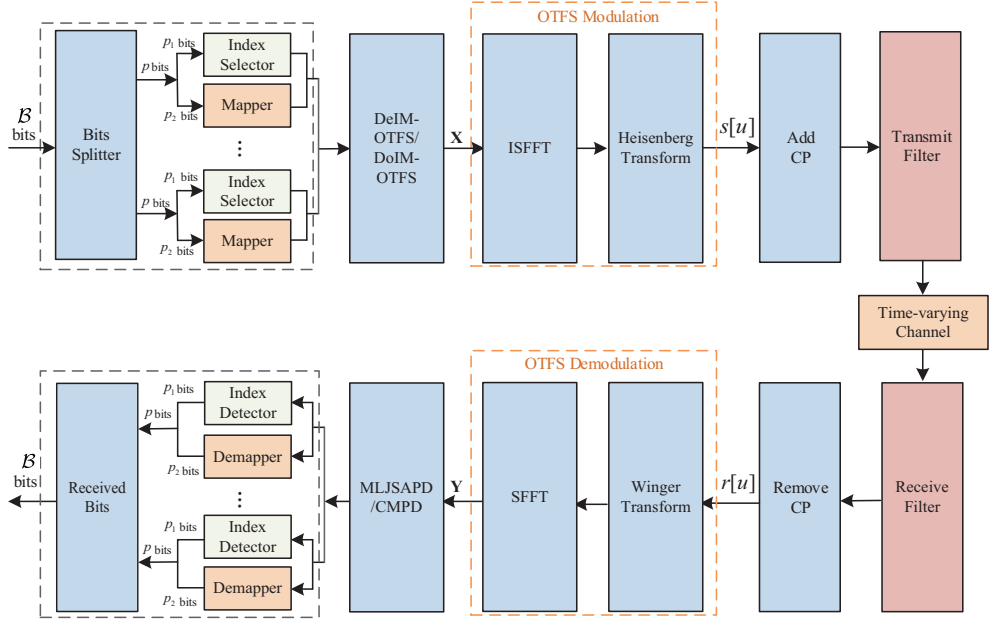}
	\caption{Transmitter and receiver structures of the proposed DeIM-OTFS/DoIM-OTFS system.}
	\label{fig:fig3}
\end{figure*}
\begin{align}
h[u, p] &=\sum_{i=1}^{L} h_{i} e^{j 2 \pi \nu_{i}\left(u T_{s}-p T_{s}\right)} {\rm{P}_{rc}}\left(p T_{s}-\tau_{i}\right), \nonumber\\
&u =0, \ldots, M N-1 ,~ p=0,\ldots, P-1,
\label{eq:3func}
\end{align}
where $h_i$, $\tau_i$ and $\nu _{i}$ denote the channel gain, delay, and Doppler shift corresponding to the $i$-th path, respectively. Parameter $L$ represents the number of multipaths. The number of the channel taps $P$ is determined by the maximal channel delay spread and the duration of the overall filter response. ${\rm{P}_{rc}}\left(p T_{s}-\tau_{i}\right)$ is the sampled overall filter response composed of bandlimiting matched filters equipped at the transmitter and receiver, which can control the bandwidth of the transmitted signal and achieve maximum signal-to-noise ratio (SNR) at the receiver. In our proposed DeIM-OTFS/DoIM-OTFS system, we choose a pair of root raised-cosine (RRC) filters in the transmitter and receiver, which are the most commonly implemented pulse shaping filters to generate an equivalent raised-cosine (RC) rolloff pulse for ${\rm{P}_{rc}}(\tau)$. Unlike the existing works in \cite{9129380,9335633}, which require delay shifts must be on the grid, we relax such ideal assumption and consider that the channel delays do not necessarily stand on the OTFS sampling grid. The Doppler frequency shift of the $i$-th path can be written as  $\nu _{i}=\left(k_{\nu_{i}}+\beta_{\nu_{i}}\right)/{NT}$,
where integer $k_{\nu_{i}}$ denotes the index of Doppler $\nu_{i}$, and real $\beta_{\nu_{i}}\in(-0.5,0.5]$ represents the fractional shift from the nearest Doppler tap $k_{\nu_{i}}$.

\subsection{Receiver Model}
At the receiver, the channel output signal first enters a receive filter. After removing the CP, the received signal can be written as
\begin{equation}
r[u]=\sum_{p=0}^{P-1} h[u, p] s\left[[u-p]_{M N}\right]+n[u], ~u=0, \ldots, M N-1,
\end{equation}
where $\textbf{n}=[n[1],n[2],\ldots,n[MN-1]]$ represents the filtered noise.

Then, the received time domain signal $\textbf{r}$ is transferred back to the time-frequency domain signal by Wigner transform (i.e., the inverse of Heisenberg transform) using a rectangular pulse $g_{r x}(t)$ at the receiver, which is given by 
\begin{align}
\overline{Y}[m, n]=& \sum_{u=0}^{M N-1} g_{r x}^{*}\left(u T_{s}-n T\right) r[u] e^{-j 2 \pi m \Delta f\left(u T_{s}-n T\right)}, \nonumber\\
& m=0, \ldots, M-1 ; n=0, \ldots, N-1.
\end{align}

Finally, the signal matrix in the time-frequency domain is processed via the symplectic finite Fourier transform (SFFT) to produce the delay-Doppler domain signal, which can be represented~as

\begin{equation}
\mathbf{Y}=\mathbf{F}_{M}^{\text{H}} \overline{\mathbf{Y}} \mathbf{F}_{N}.
\end{equation}

Based on the above analysis, the DeIM-OTFS/DoIM-OTFS input-output relationship in the delay-Doppler domain can be written as \cite{9349154}
\begin{align}
Y[\ell, k]=& \sum_{p=0}^{P-1} \sum_{i=1}^{L} \sum_{q=0}^{N-1} h_{i} {\rm{P}_{rc}}\left(p T_{s}-\tau_{i}\right) \gamma\left(k, \ell, p, q, k_{\nu_{i}}, \beta_{\nu_{i}}\right) \nonumber\\& X\left[[\ell-p]_{M},\left[k-k_{\nu_{i}}+q\right]_{N}\right]+Z[\ell, k],
\label{eq:13func}
\end{align}
where $Z[\ell,k]$ denotes the delay-Doppler domain noise sample at the output of the SFFT, and
\begin{subequations}
\begin{align}
&\gamma\left(k, \ell, p, q, k_{\nu_{i}}, \beta_{\nu_{i}}\right) \nonumber\\& = \begin{cases}\frac{1}{N} \xi\left(\ell, p, k_{\nu_{i}}, \beta_{\nu_{i}}\right) \theta\left(q, \beta_{\nu_{i}}\right),  p \leq \ell<M, \\
\frac{1}{N} \xi\left(\ell, p, k_{\nu_{i}}, \beta_{\nu_{i}}\right) \theta\left(q, \beta_{\nu_{i}}\right) \phi\left(k, q, k_{\nu_{i}}\right),  0 \leq \ell<p,\end{cases} \\
&\hspace{20mm}\xi\left(\ell, p, k_{\nu_{i}}, \beta_{\nu_{i}}\right)=e^{j 2 \pi\left(\frac{\ell-p}{M}\right)\left(\frac{k_{\nu_{i}}+\beta_{\nu_{i}}}{N}\right)}, \\
&\hspace{20mm}\theta\left(q, \beta_{\nu_{i}}\right)=\frac{e^{-j 2 \pi\left(-q-\beta_{\nu_{i}}\right)}-1}{e^{-j \frac{2 \pi}{N}\left(-q-\beta_{\nu_{i}}\right)}-1}, \\
&\hspace{20mm}\phi\left(k, q, k_{\nu_{i}}\right)=e^{-j 2 \pi \frac{\left[k-k_{\nu_{i}}+q\right]_{N}}{N}}.
\end{align}
\label{eq:8func}
\end{subequations}
\hspace{-1.5mm}We estimate the signal $\mathbf{X}$ from the received delay-Doppler signal $\mathbf{Y}$, then signal $\mathbf{X}$ is transformed into bits after a series of inverse mapping of IM. From (\ref{eq:13func}), we can observe that the off-grid Doppler shifts will spread to the whole Doppler domain, while the delay spreads only cause the ISI near the maximum delay taps. Therefore, the existing IM-OTFS works \cite{9129380} and \cite{9335633} are sensitive to inter-Doppler interference (IDI) and ISI because only individual resource unit is activated each time, leading to a performance loss. 
However, our proposed block-wise IM schemes are potentially robust to the effects of the channel. We will justify this by analyzing the performance of our proposed DeIM-OTFS/DoIM-OTFS system in the next section.

\section{Performance Analysis}
In this section, we derive BER upper bounds for the proposed DeIM-OTFS/DoIM-OTFS system, where the ML detector is used to decode the index and constellation bits.  

According to (\ref{eq:13func}), the DeIM-OTFS/DoIM-OTFS input-output relationship in the delay-Doppler domain can be vectorized as
\begin{equation}
\mathbf{y}^{\rm{T}}= \mathbf{h}\mathbf{\Phi}(\mathbf{X})+\mathbf{z}^{\rm{T}},
\end{equation}
where $\mathbf{y}^{\rm{T}}\in \mathbb{C}^{1 \times MN}$ denotes the received signal vector, $\mathbf{h}=\left [ h_{1} ,h_{2},\ldots ,h_{L}\right ]$ is a path coefficient vector and $h_{i}$ is distributed as $ \mathcal{CN}\left ( 0,1/L \right )$. $\mathbf{z}^{\rm{T}}\in \mathbb{C}^{1 \times MN}$ denotes the vector representation of \{$Z[\ell,k]$\} with $\ell=0,\ldots,M-1$ and $k=0,\ldots,N-1$. $\mathbf{\Phi}(\mathbf{X})\in \mathbb{C}^{L \times MN}$ is a signal matrix dependent on $\mathbf{X}$ whose $\rho  $-th column $(\rho  =\ell+kM , \rho  =0, \ldots, M N-1)$, denoted as $\mathbf{\Phi}_{\rho}(\mathbf{X})$, is given by (\ref{eq:16func}), as shown at the top of the next page.

\begin{figure*}[!t]
	\begin{equation}
	\mathbf{\Phi}_{\rho}(\mathbf{X})=\left[\begin{array}{c}
	\sum_{q=0}^{N-1} \sum_{p=0}^{P-1}{\rm{P}_{rc}}(pT_{s}-\tau _{1})\gamma\left(k, \ell, p, q, k_{\nu_{1}}, \beta_{\nu_{1}}\right)  X\left[[\ell-p]_{M},\left[k-k_{\nu_{1}}+q\right]_{N}\right] \\
	\sum_{q=0}^{N-1} \sum_{p=0}^{P-1}{\rm{P}_{rc}}(pT_{s}-\tau _{2})\gamma\left(k, \ell, p, q, k_{\nu_{2}}, \beta_{\nu_{2}}\right)  X\left[[\ell-p]_{M},\left[k-k_{\nu_{2}}+q\right]_{N}\right] \\
	\vdots \\
	\sum_{q=0}^{N-1} \sum_{p=0}^{P-1}{\rm{P}_{rc}}(pT_{s}-\tau _{L})\gamma\left(k, \ell, p, q, k_{\nu_{L}}, \beta_{\nu_{L}}\right)  X\left[[\ell-p]_{M},\left[k-k_{\nu_{L}}+q\right]_{N}\right]
	\end{array}\right].
	\label{eq:16func}
	\end{equation}
\end{figure*}

We assume that perfect CSI is known at the receiver. The conditional pairwise error probability (PEP) for the proposed DeIM-OTFS/DoIM-OTFS system is defined as the probability of the transmitting symbol matrix $\mathbf{X}$ and deciding $\mathbf{\widehat{X}}$, which can be given by
\begin{equation}
P(\mathbf{X} \rightarrow \mathbf{\widehat{X}} | \mathbf{h} )=Q\left(\sqrt{\frac{\|\mathbf{h}(\mathbf{\Phi}(\mathbf{X})-\mathbf{\Phi}(\mathbf{\widehat{X}})\|^{2}}{2 N_{0}}}\right).
\end{equation}

Denoting the SNR by $\gamma=1 / N_{0}$, the PEP averaged over the channel statistics is given by
\begin{equation}
P(\mathbf{X} \rightarrow \mathbf{\widehat{X}})=\mathbb{E}\left[Q\left(\sqrt{\frac{\gamma\|\mathbf{h}(\mathbf{\Phi}(\mathbf{X})-\mathbf{\Phi}(\mathbf{\widehat{X}}))\|^{2}}{2}}\right)\right],
\label{eq:18func}
\end{equation}
where,
\begin{align}
&\| \mathbf{h}( \mathbf{\Phi}(\mathbf{X})-\mathbf{\Phi}(\mathbf{\widehat{X}})  )  \|^2 \nonumber\\&=
\mathbf{h} ( \mathbf{\Phi}(\mathbf{X})-\mathbf{\Phi}(\mathbf{\widehat{X}})  ) ( \mathbf{\Phi}(\mathbf{X})-\mathbf{\Phi}(\mathbf{\widehat{X}}) )^{\text{H}}\mathbf{h}^{\text{H}}=\mathbf{h}\mathbf{\Gamma}\mathbf{h}^{\text{H}}.
\label{eq:19func}
\end{align}
Here, the matrix $\mathbf{\Gamma}$ is a Hermitian matrix that is diagonalizable by unitary transformation and it can be decomposed as $\mathbf{\Gamma}=\mathbf{U} \mathbf{\Lambda} \mathbf{U}^{H}$, where $\mathbf{U}$ is unitary and $\boldsymbol{\Lambda}=\operatorname{diag}\left\{\lambda_{1}^{2}, \ldots, \lambda_{L}^{2}\right\}$ with  $\lambda_{i}$ being the $i$-th singular value of the difference matrix $\boldsymbol{\Delta}=\mathbf{\Phi}(\mathbf{X}) -\mathbf{\Phi}(\mathbf{\widehat{X}})$. 

By defining $\tilde{\mathbf{h}}=\mathbf{h} \mathbf{U}$, we can rewrite (\ref{eq:19func}) as
\begin{align}
\| \mathbf{h}( \mathbf{\Phi}(\mathbf{X})-\mathbf{\Phi}(\mathbf{\widehat{X}})  )\|^2=\mathbf{h}\mathbf{\Gamma}\mathbf{h}^{\text{H}}=\tilde{\mathbf{h}}\mathbf{\Lambda}\tilde{\mathbf{h}}^{H}.
\label{eq:20func}
\end{align}

Therefore, (\ref{eq:18func}) can be calculated as
\begin{equation}
P(\mathbf{X} \rightarrow \mathbf{\widehat{X}})=\mathbb{E}\left[Q\left(\sqrt{\frac{\gamma \sum_{i=1}^{\alpha } \lambda_{i}^{2}\left|\tilde{h}_{i}\right|^{2}}{2}}\right)\right],
\end{equation}
where $\alpha $ denotes the rank of the difference matrix $\boldsymbol{\Delta}$ and $\tilde{h}_{i}$ is the $i$-th element of the vector $\tilde{\mathbf{h}}$. We approximate the $Q$-function quite well by using \cite{6587554}
\begin{equation}
Q\left ( x \right )\widetilde{=}\frac{1}{12}e^{-\frac{x^{2}}{2}}+\frac{1}{4}e^{-\frac{2x^{2}}{3}}.
\end{equation}
\begin{figure}
	\center
	\includegraphics[width=3.5in,height=2.4in]{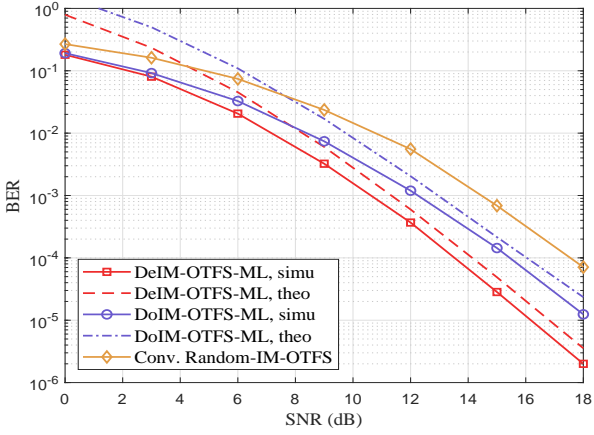}
	\caption{BER performance comparison between the proposed DeIM-OTFS/DoIM-OTFS schemes and the conventional random IM-OTFS scheme.}
	\label{fig:fig5}
\end{figure}
\hspace{-1.5mm}Then, the PEP can be approximated as
\begin{equation}
P(\mathbf{X} \rightarrow \mathbf{\widehat{X}}) \approx \frac{1}{12}\prod_{i=1}^{\alpha } \frac{1}{1+\frac{\gamma \lambda_{i}^{2}}{4 L}} + \frac{1}{4}\prod_{i=1}^{\alpha } \frac{1}{1+\frac{\gamma \lambda_{i}^{2}}{3 L}}.
\label{eq:20func}
\end{equation}
At high SNRs, (\ref{eq:20func}) can be further simplified as
\begin{equation}
\begin{aligned}
P(\mathbf{X} \rightarrow \mathbf{\widehat{X}}) \approx \frac{1/12}{\gamma^{\alpha } \underset{i=1}{\overset{\alpha }{\prod }} \frac{\lambda_{i}^{2}}{4 L}} +  \frac{1/4}{\gamma^{\alpha } \underset{i=1}{\overset{\alpha }{\prod }} \frac{\lambda_{i}^{2}}{3 L}}.
\label{eq:21func}
\end{aligned}
\end{equation}

After evaluating the unconditional PEP from (\ref{eq:21func}), the average BER of the proposed DeIM-OTFS/DoIM-OTFS scheme can be upper bounded by
\begin{equation}
P_{b}\leq\frac{1}{\mathcal{B} \varpi _{\mathbf{X}}}\sum_{\mathbf{X}}\sum_{\mathbf{\widehat{X}}}P ( \mathbf{X}\rightarrow \mathbf{\widehat{X}} )e (\mathbf{X},\mathbf{\widehat{X}}   ) ,
\end{equation}
where $\varpi_{\mathbf{X}}$ denotes the number of possible realizations of $\mathbf{X}$, and $e (\mathbf{X},\mathbf{\widehat{X}}   )$ is the number of error bits for the corresponding pairwise error event.
 \begin{figure}
	\center
	\includegraphics[width=3.5in,height=2.4in]{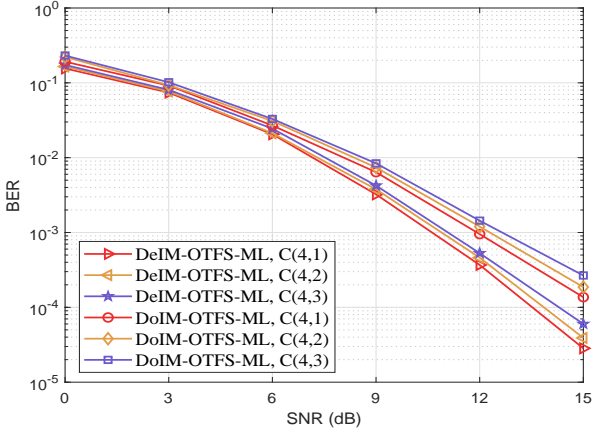}
	\caption{BER performance comparison between the proposed DeIM-OTFS and DoIM-OTFS schemes with different activation strategies.}
	\label{fig:fig6}
\end{figure}

In Fig.~\ref{fig:fig5}, we compare the BER performance of the proposed DeIM-OTFS and DoIM-OTFS schemes with that of the conventional random IM-OTFS scheme. The parameters of the considered three systems are: (i) DeIM-OTFS system with $M=N=\widehat{M}=\widehat{N}=4$, the number of active blocks is $\widehat{k}=1$, QPSK; and (ii) DoIM-OTFS system with $M=N=\widehat{M}=\widehat{N}=4$, the number of active blocks is $\widehat{k}=1$, QPSK; and (iii) random IM system with $M=N=4$, the number of active resource units is set to $2$, QPSK. All the above considered systems have the same SE of 0.625bps/Hz for fair comparison. The channel model is given by (\ref{eq:3func}), and the number of propagation paths is considered to be four (i.e., $L=4$). The velocity of mobile user is set to $\lambda=300$ Kmph and the carrier frequency is 4 GHz. As can be seen from Fig.~\ref{fig:fig5}, the simulated BER and theoretical upper bound of the DeIM-OTFS and DoIM-OTFS schemes almost coincide at the high SNR regime, which verifies the accuracy of theoretical results. 
Furthermore, our proposed DeIM-OTFS and DoIM-OTFS systems can achieve superior performance to the existing random IM-OTFS system. Moreover, the BER performance of the DeIM-OTFS scheme exhibits an approximately 2~dB gain over the DoIM-OTFS scheme under the same conditions. This is due to the fact that the effect of off-grid channel Doppler spreads causes severe interference among the resource units in the Doppler domain, while the channel delay spreads only cause interference in the delay domain with maximum delay taps. Such constrained interference of the DeIM-OTFS scheme makes the decision of the active and inactive resource blocks more accurate at the receiver than DoIM-OTFS scheme, leading to a better performance. In other words, the DoIM-OTFS scheme suffers from more severe interference than the DeIM-OTFS scheme, thus, resulting in poor performance due to the lower accuracy of receiver detection.

 \begin{figure}
	\center
	\includegraphics[width=3.5in,height=2.4in]{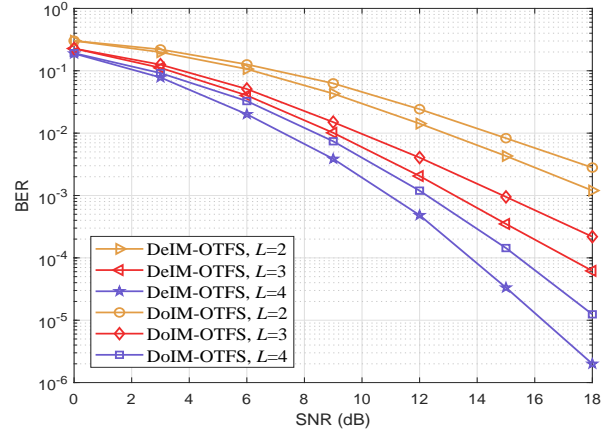}
	\caption{BER performance comparison between the proposed DeIM-OTFS and DoIM-OTFS schemes with different numbers of multipaths, where $M=N=4$, $\widehat{k}=1$ and QPSK is adopted.}
	\label{fig:fig7}
\end{figure}

Fig.~\ref{fig:fig6} gives the comparison results for the proposed DeIM-OTFS and DoIM-OTFS schemes with different SE values (0.625bps/Hz, 1.125bps/Hz and 1.625bps/Hz, respectively). In this simulation, the values of $M$, $N$, $\widehat{M}$ and $\widehat{N}$ are set to 4, and QPSK modulation is adopted. As seen from Fig.~\ref{fig:fig6}, DeIM-OTFS and DoIM-OTFS systems with index combination C(4,1) exhibit exactly the best BER performance, which means that increasing the number of active blocks will slightly decrease the BER performance of the DeIM-OTFS/DoIM-OTFS scheme. This can be understood since the detection of data symbols and active indices are more challenging for a higher SE with severe interference effect. Moreover, we again notice that the DeIM-OTFS scheme can always achieve superior BER performance to the DoIM-OTFS scheme for different activated strategies, which is consistent with the observations in Fig.~\ref{fig:fig5}.

Fig.~\ref{fig:fig7} compares the BER performance of the DeIM-OTFS and DoIM-OTFS schemes with different numbers of channel multipaths under $M=N=\widehat{M}=\widehat{N}=4$ and $\widehat{k}=1$, where all the schemes have the same SE of 0.625bps/Hz. As the number of multipaths increases from 2 to 4, we can observe a significant performance improvement in both DeIM-OTFS and DoIM-OTFS schemes. Specifically, the proposed DeIM-OTFS and DoIM-OTFS schemes of $L=4$ can achieve about 4~dB gain than that of $L=3$, while a more than 5~dB gain is obtained compared to that of $L=2$. This can be explained by the fact that with a larger number of independent resolvable multipaths, more diversity can be exploited for better performance. 
 
It is well-known that the SEs of DeIM-OTFS and DoIM-OTFS systems increase with a larger size of each subframe and higher-order signal modulation. However, these would lead to an extremely large size of the look-up table and increase the computational complexity of both the transmitter and receiver. Moreover, the computational complexity of the DeIM-OTFS and DoIM-OTFS schemes for ML detection are $\mathcal{O}((2^{p_1}  M_c^{\widehat{k}\widehat{N}})^J)$ and $\mathcal{O}((2^{p_1}  M_c^{\widehat{k}\widehat{M}})^J)$, respectively, which increase exponentially with a large size of look-up table. In order to solve this problem, we develop low-complexity MLJSAPD and CMPD algorithms for the proposed DeIM-OTFS and DoIM-OTFS systems in the next section.

\section{Receiver Design}
In this section, we develop MLJSAPD and CMPD algorithms for practical large-dimensional signal detection for the DeIM-OTFS/DoIM-OTFS system. Here, we use the DeIM-OTFS system as an example, which can be generalized to the DoIM-OTFS system in a straightforward manner. 

According to (\ref{eq:13func}), the input-output relationship of the DeIM-OTFS/DoIM-OTFS system can be vectorized as 
\begin{equation}
\mathbf{y}=\mathbf{H} \mathbf{x}+\mathbf{z},
\label{eq:15func}
\end{equation}
where $\mathbf{x}, \mathbf{y}\in \mathbb{C}^{M N \times 1}$, and $\mathbf{z} \in \mathbb{C}^{M N \times 1}$ is the noise vector. $\mathbf{H} \in \mathbb{C}^{M N \times M N}$ is a sparse matrix since the number of non-zero elements in each row and column of $\mathbf{H}$ is $\mathcal{Z}$ due to the modulo-$N$ and modulo-$M$ operations. The $( \ell+kM+1)$-th element of $\mathbf{x}$ is defined by $x[\ell+kM+1]=X[\ell, k]$ with $\ell=\widehat{M}{\overline{\ell}}+{\widehat{\ell}} \ (  0\leq \ell\leq M-1\hspace{-1mm}\ )$ and $k=\widehat{N}\overline{k}+\widehat{k} \ (0\leq k\leq N-1\hspace{-1mm}\ )$. Similarly, the $( \ell+kM+1)$-th element of $\mathbf{y}$ and $\mathbf{z}$ are $y[\ell+kM+1]=Y[\ell, k]$ and $z[\ell+kM+1]=Z[\ell, k]$, respectively, where $\ell=\widehat{M}{\overline{\ell}}+{\widehat{\ell}}$ and $k=\widehat{N}\overline{k}+\widehat{k}$. 

The joint maximum a posterior (MAP) probability detection rule of the transmitted signal is given by
\begin{equation}
\widehat{\mathbf{x}}=\underset{\mathbf{x} \in \{S \cup 0\}^{MN \times 1}}{\arg \max } \operatorname{Pr}(\mathbf{x} | \mathbf{y},\mathbf{H}),
\label{eq:22func}
\end{equation}
where ``0'' means the resource units is not activated; otherwise it is activated. 

We observe that the exact computation of (\ref{eq:22func}) has a complexity exponential in $MN$, making the joint MAP detection intractable for practical values of $N$ and $M$. To reduce the receiver complexity, we propose two efficient detection algorithms in the following subsections.

\subsection{MLJSAPD Algorithm}
In this subsection, the details of the MLJSAPD algorithm are described in the following and summarized in $\textbf {Algorithm 1}$.

\begin{figure}
	\centering
	\includegraphics[width=3.5in,height=2.3in]{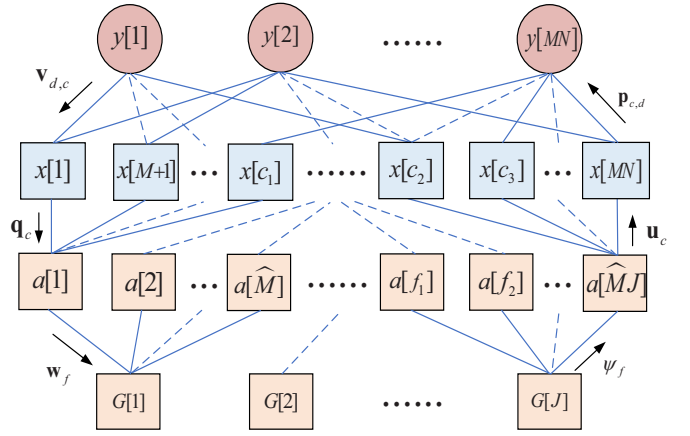}
	\caption{Graphical model for message passing in the proposed MLJSAPD algorithm, where $c_1=(\widehat{N}-1)M+1$, $c_2=M(N-\widehat{N}+1)$, $c_3=M(N-\widehat{N}+2)$, $f_1=\widehat{M}(J-1)+1$ and $f_2=\widehat{M}(J-1)+2$.}
	\label{fig:fig4}
\end{figure}

In (\ref{eq:22func}), $ \operatorname{Pr}(\mathbf{x}|\mathbf{y},\mathbf{H})$ can be written as:
\begin{align}
&\operatorname{Pr}\left ( \mathbf{x,a| y,\mathbf{H} }\right )\nonumber\\ &\propto \operatorname{Pr}\left ( \mathbf{y| x,a,\mathbf{H} }\right )\operatorname{Pr}\left ( \mathbf{x,a}\right )\nonumber\\
&=\operatorname{Pr}\left ( \mathbf{y| x,a ,\mathbf{H}}\right )\operatorname{Pr}\left ( \mathbf{x| a }\right )\operatorname{Pr}\left ( \mathbf{ a }\right ) \nonumber\\
&=\hspace{-0.5mm}\hspace{-0.5mm} \prod_{d=1 }^{MN}\hspace{-1mm}\operatorname{Pr}\hspace{-0.5mm}\left ( y[d]| \mathbf{x,a,\mathbf{H}} \right ) \hspace{-1mm}\prod_{f=1 }^{\widehat{M}J} \prod_{c=\overline{k}(\widehat{N}-1)M+f}^{(\overline{k}+1)(\widehat{N}-1)M+f}\hspace{-2mm}\operatorname{Pr}\left ( x[c]| a[f] \right )\hspace{-0.5mm}\operatorname{Pr}\left ( \mathbf{a} \right ),
\label{eq:25func}
\end{align}
where $\mathbf{a}=[a[1],a[2],\ldots,a[\widehat{M}J]]$ denotes the vector of the activation state of all blocks. We assume the components of $\mathbf{y}$ are approximately independent for a given $\mathbf{x}$, and any symbols $\mathbf{ x } \in \{S \cup 0\}$ are transmitted with equal probability. 

As shown in Fig.~\ref{fig:fig4}, we interpret the system model as a sparsely connected factor graph with four types of nodes: $(i)~ MN $ observation nodes corresponding to the elements of $\mathbf{y}$, $(ii)~MN$ variable nodes corresponding to the elements of $\mathbf{x}$, $(iii) $ $\widehat{M}J$ activity indicator nodes corresponding to the elements of $\mathbf{ a }$, $(iv)$ $J$ constraint nodes $G$. Let $\mathcal{I}(d)$ and $\mathcal{J}(c)$ denote the sets of indexes with nonzero elements in the $d$-th row and $c$-th column of $\mathbf{H}$, where $d=1, \ldots,   MN$ and $c=1, \ldots,  MN$. Each observation node $y[d]$ is connected to the set of $\mathcal{Z}$ variable nodes $\{x[c], c \in \mathcal{I}(d)\}$ while each variable node $x[c]$ is connected to the set of $\mathcal{Z}$ observation nodes $\{y[d], d \in \mathcal{J}(c)\}$. Specifically, from constraint node to indicator node, each constraint node $G[\beta]~(1\leq \beta \leq J)$ is connected to the set of $\widehat{M}$ indicator nodes $a[f]$, where $f\in\mathcal{K}(\beta)$ and $\mathcal{K}(\beta)= [(\beta -1)\widehat{M}+1,\ldots ,\beta\widehat{M}]$. From indicator node to variable node, each indicator node $a[f]~( 1\leq f \leq \widehat{M}J)$ is connected to the set of $\widehat{N}$ variable nodes $x[c]$ with parameter $c\in\mathcal{D}(f)$ and $\mathcal{D}(f)= [\overline{k}(\widehat{N}-1)M+f,\overline{k}(\widehat{N}-1)M+f+M,\ldots ,(\overline{k}+1)(\widehat{N}-1)M+f]$. Note that observation nodes and variable nodes denote Layer 1 which generates approximate posterior probabilities of the individual elements $\mathbf{ x }$. The variable nodes and activity indicator nodes denote Layer 2, which generates the probability estimate of message from $x[c]$ to $a[f]$. Furthermore, activity indicator nodes and constraint nodes denote Layer 3, which generates approximate probabilities of the individual elements $\mathbf{ a }$ being active or inactive. For the proposed MLJSAPD algorithm, its detailed steps in iteration $n_{iter}$ are described below. 

\begin{algorithm}[t]
	\setstretch{1.2}
	\caption{Proposed MLJSAPD Algorithm}\label{alg:alg1}
	\begin{algorithmic}
		\STATE
		\STATE {\text{Input:}} $\mathbf{y}, \mathbf{H},\sigma^{2}$ and $n_{iter}^{max}$.
		\STATE {\text{Initialization:}} $p_{c,d}\left ( x \right )=\frac{1}{|S\cup 0|},q_{c}\left ( 1 \right )=q_{c}\left ( 0\right )= \frac{1}{2},\psi _{f}\left ( 1 \right )=\psi _{f}\left ( 0\right )= \frac{1}{2}~\forall c,d,f$, $\eta ^{(0)}=0$ and iteration count $n_{iter}=1$.
		\STATE $\textbf{repeat}$
		\STATE \hspace{0.35cm}1) Each observation node $y[d]$ generates the calculated mean $\mu_{d, c}^{n_{iter}}$ and variance $(\sigma_{d, c}^{n_{iter}})^{2}$, given in (\ref{eq:27func}) and (\ref{eq:28func}), then send $\mathbf{v}_{d,c}^{n_{iter}}$ to the connected variable nodes $x[c]$;
		\STATE \hspace{0.35cm}2) Each variable node $x[c]$ generates $\mathbf{ q }_{c}^{n_{iter}}$ in (\ref{eq:30func}) and transmits them to the connected indicator node $a[f]$;		
		\STATE \hspace{0.35cm}3) Each indicator node $a[f]$ computes $\mathbf{ w }_{f}^{n_{iter}}$ in (\ref{eq:32func}) and then transmits them to the connected constraint node $G[\beta]$;
		\STATE \hspace{0.35cm}4) Each constraint node $G[\beta]$ generates $\bm{ \psi }_{f}^{n_{iter}}$ in (\ref{eq:33func}), and sends them to the connected indicator nodes $a[f]$;
		\STATE \hspace{0.35cm}5) Each indicator node $a[f]$ computes $\mathbf{ u }_{c}^{n_{iter}}$ in (\ref{eq:34func}), and delivers them to the connected variable nodes $x[c]$;
		\STATE \hspace{0.35cm}6) Each variable node $x[c]$ computes $\mathbf{p}_{c,d}^{n_{iter}}$ in (\ref{eq:36func}), and delivers them to the connected variable nodes $y[d]$;
		\STATE \hspace{0.35cm}7) Compute the convergence indicator $\eta ^{n_{iter}}$ and the probability of the transmitted symbols $\mathbf{ p }_{c}^{n_{iter}}$ in (\ref{eq:37func}) and (\ref{eq:40func}), respectively;
		\STATE \hspace{0.35cm}8) Update the symbol probabilities $\overline{\mathbf{ p }}_{c}=\mathbf{ p }_{c}^{n_{iter}}$ if $\eta ^{n_{iter}}>\eta ^{n_{iter}-1}$;
		\STATE \hspace{0.35cm}9) $n_{iter}= n_{iter}+1$;
		\STATE {\textbf{until}} $\eta ^{n_{iter}}=1$ or $n_{iter}=n_{iter}^{max}$.
		\STATE Output: The decisions of the transmitted symbols ${\widehat x}\left [ c \right ],~c\in \mathbb{A}$.
	\end{algorithmic}
	\label{alg1}
\end{algorithm}

$\textbf{1) From observation node}$ $ y[d]$ $\textbf{to variable nodes}$ $ x[c],$ $c\in \mathcal{I} (d)$:
At each observation node, we calculate the extrinsic messages for each connected variable node according to the sparsity channel model, and prior information from other connected variable nodes. The interference is approximately modeled as a Gaussian random variable $\zeta_{d, c}^{n_{iter}}$, where $\mu_{d, c}^{n_{iter}}$ and variance $(\sigma_{d, c}^{n_{iter}})^{2}$ denote the mean and variance, respectively. Thus, the received signal $y[d]$ can be written as
\begin{equation}
y[d]=x[c] H[d, c]+\underbrace{\sum_{e \in \mathcal{I}(d), e \neq c} x[e] H[d, e]+v[d]}_{\zeta_{d, c}^{n_{iter}}},
\label{eq:23func}
\end{equation}
with
\begin{equation}
\mu_{d, c}^{n_{iter}}=\sum_{e \in \mathcal{I}(d), e \neq c} H[d, e]\sum_{x\in \{S \cup 0\}} p_{e, d}^{n_{iter}-1}\left(x\right) x ,
\label{eq:27func}
\end{equation}
and
\begin{align}
(\sigma_{d, c}^{n_{iter}})^{2}&= \sum_{e \in \mathcal{I}(d), e \neq c}\left(\sum_{x\in \{S \cup 0\}} p_{e, d}^{n_{iter}-1}\left(x\right)\left|x\right|^{2}|H[d, e]|^{2}\right.
\nonumber\\& \left.-\left|\sum_{x\in \{S \cup 0\}} p_{e, d}^{n_{iter}-1}\left(x\right) x H[d, e]\right|^{2}\right)+\sigma^{2} ,
\label{eq:28func}
\end{align}
where $\sigma^{2}=\sigma _{N}^{2}\int _{\mu }{\rm{P}_{rrc}^{2}}(\mu )d\mu $ is the variance of the colored Gaussian noise. ${\rm{P}_{rrc}}(\mu )$ denotes the RRC rolloff receive filter and $\sigma _{N}^{2}$ is the variance of the AWGN at the receiver input \cite{9349154}. The mean $\mu _{d,c}^{n_{iter}}$ and variance $(\sigma_{d,c}^{n_{iter}})^{2}$ of the interference terms are used to calculate the approximate marginal probability of the transmitted symbols. Therefore, the probability estimate of $x[c]$ passed from observation node $y[d]$ to variable node $x[c]$ is given by
\begin{align}
&v_{d,c}^{n_{iter}}(x)\propto{\rm{ Pr }}(y[d]|x[c]=x,\mathbf{H})\nonumber\\
&\propto \exp \left (-\frac{ \left | y[d]-\mu _{d,c}^{n_{iter}}-H[d,c]x \right |^{2}}{(\sigma  _{d,c}^{n_{iter}})^{2}}\right ),~\forall x\in \{S \cup 0\}.
\label{eq:26func}
\end{align}

$\textbf {2) From variable node}$ $x[c]$ $\textbf {to activity indicator node}$ $a[f]$: All resource units in each block are connected to an indicator node $a[f]$. The probability of each indicator node $a[f]$ is determined by the corresponding variable nodes. We assume the probability estimate of message from $x[c]$ to $a[f]$ is given by
\begin{equation}
\begin{aligned}
q_{c}^{n_{iter}}\left ( b \right )=\Delta \cdot \widetilde{q}_{c}^{n_{iter}}\left ( b \right )+\left ( 1-\Delta  \right )\cdot q_{c}^{n_{iter-1}}\left ( b \right ),
\end{aligned}
\label{eq:30func}
\end{equation}
where $\Delta \in(0,1]$ is the $message$ $damping$ $factor$ used to improve the system performance by controlling the convergence rate, and
\begin{subequations}
\begin{align}
\widetilde{q}_{c}^{n_{iter}}\left ( b \right )&\overset{\Delta }{=}{\rm{ Pr }}\left ( a[f]=b|\mathbf{x} \right )\\&\propto\left\{\begin{matrix}
\underset{x\in S }{\sum }~\underset{d\in \mathcal{J}(c)}{\prod }{\rm{ Pr }}\left ( y[d]|x[c]=x,\mathbf{H} \right ),~&{\rm{if}} ~b=1,\\ 
\hspace{-6mm}\underset{d\in \mathcal{J}(c)}{\prod }{\rm{ Pr }}\left ( y[d]|x[c]=0,\mathbf{H} \right ),~&{\rm{if}} ~b=0,
\end{matrix}\right.\\
&\propto\left\{\begin{matrix}
\underset{x\in S }{\sum }~\underset{d\in \mathcal{J}(c)}{\prod }v_{d,c}^{n_{iter}}\left ( x \right ),~&{\rm{if}}~b=1,\\ 
\hspace{-6mm}\underset{d\in \mathcal{J}(c)}{\prod }v_{d,c}^{n_{iter}}\left ( 0 \right ),~&{\rm{if}}~b=0.
\end{matrix}\right.
\end{align}
\label{eq:31func}
\end{subequations}

$\textbf {3) From activity indicator node}$ $a[f]$ $\textbf {to constraint node}$ $G[\beta]$: According to the indices of the activated resource units, the probability estimate of message passed from indicator node $a[f]$ to constraint node $G[\beta]$, can be written as
\begin{subequations}
\begin{align}
w_{f}^{n_{iter}}\left ( b \right )&={\rm{ Pr }}\left ( a[f] =b|\mathbf{x}^{f}\right )\\&\propto\left\{\begin{matrix}
\underset{c\in\mathcal{D}(f)}{\prod }q_{c}^{n_{iter}}\left ( 1 \right ),~{\rm{if}}~b=1,\\ 
\underset{c\in\mathcal{D}(f)}{\prod }q_{c}^{n_{iter}}\left ( 0 \right ),~{\rm{if}}~b=0,
\end{matrix}\right.
\end{align}
\label{eq:32func}
\end{subequations}
\hspace{-2mm}where $\mathbf{x}^{f}\hspace{-1mm}=\hspace{-0.5mm}[ x[\overline{k}(\widehat{N}-1)M+f],x[\overline{k}(\widehat{N}-1)M+f+M],\ldots ,x[(\overline{k}+1)(\widehat{N}-1)M+f ] ]$.

$\textbf {4) From constraint node}$ $G[\beta] \hspace{0.5mm}\textbf{to activity indicator nodes}$ $a[f]$, $f\in\mathcal{K}(\beta)$: In each subframe,
$\widehat{M}$ indicator nodes $a[f]$ are linked to a constraint node $G[\beta]$, where $\sum_{f=(\beta-1)\widehat{M}+1}^{\widehat{M}\beta}a[f]=\widehat{k}$. At each constraint node, the extrinsic information for each indicator node can be generated by prior messages collected from other indicator nodes. We can calculate the probability estimate of message passed from constraint node to indicator node, as 
\begin{subequations}
\begin{align}
&\psi_{f}^{n_{iter}}\left ( b \right )={\rm{ Pr }}\left ( a[f] =b|\mathbf{a}^{\beta}_{\setminus f}\right )\\&=\left\{\begin{matrix}
{\rm{ Pr }}\left ( \underset{e=\widehat{M}(\beta-1)+1,e\neq f}{\overset{\widehat{M}\beta}{\sum }} a[e]=\widehat{k}-1|\mathbf{a}_{\setminus f}^{\beta}\right ),~&{\rm{if}}~b=1,\\ 
\hspace{-5mm}{\rm{ Pr }}\left ( \underset{e=\widehat{M}(\beta-1)+1,e\neq f}{\overset{\widehat{M}\beta}{\sum }} a[e]=\widehat{k}|\mathbf{a}_{\setminus f}^{\beta}\right ),~&{\rm{if}}~b=0,
\end{matrix}\right.\\
&\approx \left\{\begin{matrix}
\Omega_{f}^{n_{iter}} ( \widehat{k}-1  ),~&{\rm{if}}~b=1,\\ 
\hspace{-6mm}\Omega_{f} ^{n_{iter}} ( \widehat{k}  ),~&{\rm{if}}~b=0,
\end{matrix}\right.
\end{align}
\label{eq:33func}
\end{subequations}
\hspace{-2mm}where $\mathbf{a}_{\setminus f}^{\beta}$ denotes $\mathbf{a}^{\beta}$ excluding $a[f]$ for $f\in\mathcal{K}(\beta)$ and $\mathbf{a}^{\beta}=[a[\widehat{M}(\beta-1)+1],a[\widehat{M}(\beta-1)+2],\ldots ,a[\widehat{M}\beta]]$. $\bm{\Omega } _{f}^{n_{iter}}$ is calculated as $\bm{\Omega} _{f}^{n_{iter}} = \otimes_{e=\widehat{M}(\beta-1)+1,e\neq f}^{\widehat{M}\beta}\mathbf{ w}_{e}^{n_{iter}} $ for  $\mathbf{ w}_{e}^{n_{iter}}=[w_e^{n_{iter}}(0)~ w_e^{n_{iter}}(1)]$, where $\otimes$ denotes the convolution operator.

$\textbf {5) From activity indicator node}$ $a[f]$ $\textbf {to variable nodes}$ $x[c]$, $c\in\mathcal{D}(f)$: We note that the indicator node $a[f]=1$ only when all variable nodes $x[c]$ connected to $a[f]$ are activated (i.e., $x[c]=x\in S$) and $0$ otherwise. The probability estimate of message passed from $a[f]$ to $x[c]$ is given by
\begin{subequations}
\begin{align}
u_{c}^{n_{iter}}\left ( b \right )&\overset{\Delta }{=}{\rm{ Pr }}\left ( a[f]=b|\mathbf{ x }_{\setminus c}^{f} \right )\\
&=\left\{\begin{matrix}
{\rm{ Pr }}\left ( {\underset{e\in\mathcal{D}(f),e\neq c}{\sum }}x[e]\neq 0|\mathbf{ x }_{\setminus c}^{f}\right ),~{\rm{if}} ~b =1,\\ 
\hspace{1mm}{\rm{ Pr }}\left ( {\underset{e\in\mathcal{D}(f),e\neq c}{\sum }}x[e]=0|\mathbf{ x }_{\setminus c}^{f}\right ),~{\rm{if}} ~b =0,
\end{matrix}\right.\\
&\approx \left\{\begin{matrix}
\underset{e\in\mathcal{D}(f),e\neq c}{\sum }\psi_{f}^{n_{iter}}\left ( 1 \right )q_{e}^{n_{iter}}\left ( 1 \right ),~{\rm{if}}~b=1,\\ 
\underset{e\in\mathcal{D}(f),e\neq c}{\sum }\psi_{f}^{n_{iter}}\left ( 0 \right )q_{e}^{n_{iter}}\left ( 0 \right ),~{\rm{if}}~b=0,
\end{matrix}\right.
 \end{align}
 \label{eq:34func}
\end{subequations}
\hspace{-2mm}where $\mathbf{ x }_{\setminus c}^{f}$ denotes $\mathbf{x}^{f}$ excluding $x[c]$ for $c\in\mathcal{D}(f)$. $\psi_{f}^{n_{iter}}( 1 )$ denotes the activated probability of all variable nodes connected to $a[f]$.

$\textbf {6) From variable node}$ $x[c]$ $\textbf {to observation nodes} $ $y[d],d\in \mathcal{J} (c)$: The posterior probability of the elements $\mathbf{x}$ passed from variable node $x[e]$ to observation node $y[d]$ is denoted by $\mathbf{p}_{e,d}$. At each variable node, the extrinsic information for each connected observation node is generated from prior messages collected from other observation nodes and indicator nodes. Hence, the probability $\widetilde{\mathbf{p}}_{c,d}^{n_{iter}}$ can be given by
\begin{subequations}
\begin{align}
\widetilde{p}_{c,d}^{n_{iter}}(x)&\propto u_{c}^{n_{iter}} ( x^{\bigodot } ) \prod_{e\in \mathcal{J}(c),e\neq d}{\rm{ Pr }}\left ( y[e]|x[c]=x,\mathbf{H} \right )\\&\propto u_{c}^{n_{iter}} ( x^{\bigodot } )\prod_{e\in \mathcal{J}(c),e\neq d}v_{e,c}^{n_{iter}}(x),~\forall x\in \{S \cup 0\},
\label{eq:35func}
\end{align}
\end{subequations}
where $x$$^{\bigodot }=1$ if $x\in S$ or 0 otherwise. The message from variable node $x[c]$ to observation node $y[d]$ contains the probability mass function (pmf) with elements 
\begin{equation}
\begin{aligned}
p_{c,d}^{n_{iter}}\left ( x \right )=\Delta \cdot \widetilde{p}_{c,d}^{n_{iter}}(x)+\left ( 1-\Delta  \right )\cdot p_{c,d}^{n_{iter-1}}\left ( x \right ).
\label{eq:36func}
\end{aligned}
\end{equation}

$\textbf{7) Convergence ~indicator }$: We calculate the convergence indicator $\eta ^{n_{iter}}$ for some small $\varrho$ as 
\begin{equation}
\begin{aligned}
\eta ^{n_{iter}}=\frac{1}{MN}\sum_{c=1}^{MN}\mathbb{I}\left ( \underset{x\in \{S \cup 0\}}{{\text{max}}} ~p_{c}^{n_{iter}}\left ( x \right )\geq 1-\varrho \right ),
\label{eq:37func}
\end{aligned}
\end{equation}
where $\mathbb{I}$ denotes indicator function. The posterior probability for each element of the transmit symbol is given by 
\begin{align}
p_{c}^{n_{iter}}\left ( x\right )=\frac{1}{C}u_{c}^{n_{iter}}\left ( x^{\odot } \right )\underset{d\in \mathcal{J}(c) }{\prod }v_{d,c}^{n_{iter}}\left ( x \right ),~\forall x\in \{S \cup 0\},
\label{eq:40func}
\end{align}
where $C$ is a normalizing constant.

$\textbf{8) Update~ criteria}$: If $\eta ^{n_{iter}}>\eta ^{n_{iter}-1}$, the probability of the transmitted symbols is updated only when the current iteration provides a better solution than the previous one,
\begin{align}
\overline{\mathbf{ p }}_{c}=\mathbf{ p }_{c}^{n_{iter}},~ c=1,\ldots ,MN.
\end{align}

In this algorithm, the different messages passed in this graph are as follows: $\mathbf{v}_{d,c}$ passes from observation node $y[d]$ to the connected variable node~$x[c]$; $\mathbf{ q }_c$ passes from variable node $x[c]$ to the connected activity indicator node~$a[f]$; $\mathbf{ w }_f$ passes from activity indicator node $a[f]$ to the connected constraint node~$G[\beta]$; $\bm{ \psi }_f$ passes from constraint node $G[\beta]$ to the connected activity indicator node~$a[f]$; $\mathbf{ u }_c$ passes from activity indicator node $a[f]$ to the connected variable node~$x[c]$; $\mathbf{p}_{c,d}$ passes from variable node $x[c]$ to the connected observation node~$y[d]$. All of the messages are exchanged between these four nodes until convergence.

$\textbf{9) Stopping~ criteria}$: The MLJSAPD algorithm stops when $\eta ^{n_{iter}}=1$ or the maximum number of iterations $n_{iter}^{max}$ is reached. 

After satisfy the convergence of the algorithm, we can find all of the activated blocks, which are determined by choosing the blocks of the corresponding $\widehat{k}$ largest activated probabilities in each subframe, given by $[a_{1}^{\beta },a_{2}^{\beta },a_{t}^{\beta },\ldots,a_{\widehat{k}}^{\beta }]$ with $1\leq t\leq \widehat{k}$ and $1\leq \beta \leq J$. Let $[d_{1}^{\beta },d_{2}^{\beta },\ldots,d_{\widehat{M}}^{\beta }]$ denotes the blocks of the $\beta$-th subframe, i.e., $a_{t}^{\beta }\in [d_{1}^{\beta },d_{2}^{\beta },\ldots,d_{\widehat{M}}^{\beta }]$. Then, we can obtain the corresponding activated resource units according to the activated blocks. 

Finally, we make a decision of the transmitted symbols, as given by
\begin{align}
{\widehat x}\left [ c \right ]=\underset{x\in S  } {\text{argmax}} ~\overline{p}_{c}\left ( x \right ), ~c\in \mathbb{A},
\label{eq:41func}
\end{align}
where $\mathbb{A}$ denotes the set of active resource units. 

According to the active resource units, the corresponding symbols will be transferred into bits through a series of inverse mapping of IM.
\begin{algorithm}[t]
	\setstretch{1.2}
	\caption{Proposed CMPD Algorithm}\label{alg:alg2}
	\begin{algorithmic}
		\STATE
		\STATE {\text{Input:}} $\mathbf{y}, \mathbf{H},\sigma^{2}$ and $n_{iter}^{max}$.
		\STATE {\text{Initialization:}} $p_{c,d}\left ( x \right )\hspace{-0.3mm}=\hspace{-0.3mm}\frac{1}{|S\cup 0|},\forall c,d$, $\eta ^{(0)}\hspace{-0.3mm}=\hspace{-0.3mm}0$ and $n_{iter}\hspace{-0.3mm}=\hspace{-0.3mm}1$.
		\STATE $\textbf{repeat}$
		\STATE \hspace{0.35cm}1) Each observation node $y[d]$ generates the mean $\mu_{d, c}^{n_{iter}}$ and variance $(\sigma_{d, c}^{n_{iter}})^{2}$, given in (\ref{eq:27func}) and (\ref{eq:28func}), respectively, and then transmits $\mathbf{v}_{d,c}^{n_{iter}}$ to the connected variable nodes $x[c]$;
		\STATE \hspace{0.35cm}2) Each variable node $x[c]$ computes $\mathbf{ p }_{c,d}^{n_{iter}}$ in (\ref{eq:52func}), and delivers them to the connected observation nodes $y[d]$;
		\STATE \hspace{0.35cm}3) Compute the convergence indicator $\eta ^{n_{iter}}$ and the probability of the transmitted symbols $\mathbf{ p }_{c}^{n_{iter}}$;
		\STATE \hspace{0.35cm}4) Update the symbol probabilities $\overline{\mathbf{ p }}_{c}=\mathbf{ p }_{c}^{n_{iter}}$ if $\eta ^{n_{iter}}>\eta ^{n_{iter}-1}$;
		\STATE \hspace{0.35cm}5) $n_{iter}= n_{iter}+1$;
		\STATE {\textbf{until}} $\eta ^{n_{iter}}=1$ or $n_{iter}=n_{iter}^{max}$.
		\STATE Output: The decisions of the transmitted symbols in (\ref{eq:41func}).
	\end{algorithmic}
	\label{alg2}
\end{algorithm}

\subsection{CMPD Algorithm}
To further reduce the complexity, we propose the CMPD algorithm to simplify the structure of the above factor graph and only keep the observation node $\mathbf{y}$ and variable node $\mathbf{x}$. In our proposed CMPD algorithm, we identify the active blocks by comparing the LLRs of each block after the iterations. The details of the CMPD algorithm are given as follows and summarized in $\textbf {Algorithm 2}$. 

The joint MAP probability of the transmitted signal is give by (\ref{eq:22func}).
Differently, we calculate (\ref{eq:22func}) by the following approximation:
\begin{align}
{\widehat x}\left [ c \right ]&=\underset{x\in \{S \cup 0\}}{\arg \max } \operatorname{Pr}(x[c]=x | \mathbf{y},\mathbf{H})\nonumber\\
&\propto \underset{x\in \{S \cup 0\}}{\arg \max }\prod_{d\in \mathcal{J}(c)}{\rm{ Pr }}\left ( y[d] | x[c]=x,\mathbf{H}\right ).
\end{align}
Similar to the MLJSAPD algorithm, we employ the Gaussian approximation to the interference term, and the received signal can be obtained by applying the same expression in (\ref{eq:23func}). The mean and variance of the interference in the $n_{iter}$-th iteration can still be given in (\ref{eq:27func}) and (\ref{eq:28func}), respectively. The probability estimate of $x[c]$ passed from observation node $y[d]$ to variable node $x[c]$ is given by (\ref{eq:26func}). From variable node $x[c]$ to observation node $y[d]$, the pmf vector $\mathbf{ p }_{c,d}$ is updated by the expression (\ref{eq:36func}), with
\begin{align}
\widetilde{p}_{c,d}^{n_{iter}}(x)&\propto \prod_{e\in \mathcal{J}(c),e\neq d}{\rm{ Pr }}\left ( y[e] | x[c]=x,\mathbf{H}\right )\nonumber\\
&=\prod_{e\in \mathcal{J}(c),e\neq d}\frac{v_{e,c}^{n_{iter}}(x)}{\underset{x\in \{S \cup 0\}}{\sum }v_{e,c}^{n_{iter}}(x)}.
\label{eq:52func}
\end{align}
Here, we calculate the convergence indicator $\eta ^{n_{iter}}$ by (\ref{eq:37func}). The posterior probability for each element of the transmit symbol is given as
\begin{align}
p_{c}^{n_{iter}}\left ( x\right )=\prod_{e\in \mathcal{J}(c)}\frac{v_{e,c}^{n_{iter}}(x)}{\underset{x\in \{S \cup 0\}}{\sum }v_{e,c}^{n_{iter}}(x)},~\forall x\in \{S \cup 0\}.
\label{eq:55func}
\end{align}

The update and stopping criteria of the CMPD algorithm is the same as that of the MLJSAPD algorithm. Once the stopping criteria is satisfied, we can obtain the LLR of each resource unit,~as
\begin{align}
{\widehat{ L }}[c]=\log\frac{\underset{x\in S}{\prod }p_{c}^{n_{iter}}(x)}{p_{c}^{n_{iter}}(x=0)},~ c=1,\ldots ,MN.
\label{eq:54func}
\end{align}
Then, we average the LLRs of all resource units in each block. The active blocks can be determined by choosing the blocks of the corresponding $\widehat{k}$ largest average LLRs.

Finally, we make the decisions of the transmitted symbols ${\widehat x}\left [ c \right ]$ for the active resource units according to (\ref{eq:41func}). Then, the estimated symbols are transferred into bits by using a series of inverse mapping of IM.

\subsection{Complexity Analysis}
The complexity of the proposed MLJSAPD and CMPD algorithms are analyzed in this subsection. We take the DeIM-OTFS scheme as an example, the complexity of the DoIM-OTFS scheme can be generated in a straightforward manner. As shown in TABLE II, the complexity of the proposed MLJSAPD and CMPD algorithms for each iteration is calculated according to the real-field multiplications \cite{9354639}, and exponential functions, respectively, given at the top of next page. Complex multiplication, inverse, and division are equivalent to three, four, and six real-field multiplications, respectively. The MLJSAPD algorithm complexity is mainly dominated by (\ref{eq:27func})-(26), and (28)-(32). The number of real-filed multiplications required in steps (\ref{eq:27func}), (\ref{eq:28func}), and (28)-(32) are $2MN\mathcal{Z}(M_c+1)$, $MN\mathcal{Z}(4(M_c+1)+1)$, $MN\mathcal{Z}(M_c+1)$, $2\widehat{M}J\mathcal{Z}$, ${\widehat{M}}^2{J}^2+\widehat{M}J$, $2J\mathcal{Z}$ and $MN{\mathcal{Z}}^2$, respectively. In addition, (26) is a exponential function with the complexity of $MN\mathcal{Z}$. The CMPD algorithm complexity is dominated by (\ref{eq:27func})-(26), (39), and (40). The number of real-filed multiplications of (\ref{eq:27func})-(26) is the same as the MLJSAPD algorithm, and (39), (40) are given by $MN(\mathcal{Z}+M_c+4)$ and $MN(\mathcal{Z}+M_c+8)$, respectively. From these analysis, we can observe that our proposed MLJSAPD and CMPD algorithms have tolerable complexity for symbol detection. Moreover, simulation results verified the desired performance of the MLJSAPD and CMPD algorithms in the next section.
 
\begin{table*}[!t]
	\caption{Complexity comparison of the proposed MLJSAPD and CMPD algorithms for DeIM-OTFS scheme in each iteration. \label{tab:table2}}
	\centering
	\begin{tabular}{p{2.5cm}<{\centering}|p{9.5cm}<{\centering}|p{2.5cm}<{\centering}}
		\hline
		\text{Algorithm} & \text{Real-field Multiplication}& \text{Exponential}\\
		\hline
		MLJSAPD & $MN\mathcal{Z}(8(M_c+1)+\mathcal{Z}+1)+\widehat{M}J(2\mathcal{Z}+\widehat{M}J+1)+2J\mathcal{Z}$& $MN\mathcal{Z}$\\
		\hline
		CMPD & $MN\mathcal{Z}(6(M_c+1)+1)+MN(2\mathcal{Z}+2M_c+12)$& $MN\mathcal{Z}$\\
		\hline
	\end{tabular}
\end{table*}
\begin{table}[!t]
	\caption{Simulation Parameters. \label{tab:table3}}
	\centering
	\begin{tabular}{|p{5cm}<{\centering}|p{3cm}<{\centering}|}
		\hline
		\text{Parameter} & \text{Value}\\
		\hline
		\hline
		Carrier frequency & 4 GHz\\
		\hline
		No. of subcarriers ($M$) & 64\\
		\hline
		No. of OTFS symbols ($N$) & 16\\
		\hline
		Subcarrier spacing & 15 kHz\\
		\hline
		Modulation alphabet & BPSK, QPSK\\
		\hline
		UE speed & 300, 500 Kmph\\
		\hline
	\end{tabular}
\end{table}

\section{Simulation Results}
In this section, we study the BER performance of the proposed DeIM-OTFS/DoIM-OTFS scheme with MLJSAPD and CMPD detection algorithms. We assume that the perfect channel knowledge is known at the receiver and all relevant simulation parameters are given in Table III. We also test the receiver performance of the proposed schemes with imperfect CSI.
The Doppler frequency shift of the $i$-th channel path is generated by $\nu _{i}=\nu _{max}\text{cos}(\theta_i)$, where $\nu _{max}$ denotes the maximum Doppler frequency shift and $-\pi\leq\theta_i\leq\pi$. Moreover, the RRC rolloff factor is set to 0.4 at the transmitter and receiver. Without loss of generality, we choose $\Delta=0.4$ and $\varrho =0.1$. Unless otherwise mentioned, the numbers of delay bins, Doppler bins and active blocks in each subframe are set to $\widehat{M}=4$, $\widehat{N}=4$ and $\widehat{k}=1$, respectively. The number of multipaths is set to $L=4$, and the user velocity is set to 300 Kmph. 

\begin{figure}
	\center
	\includegraphics[width=3.5in,height=2.4in]{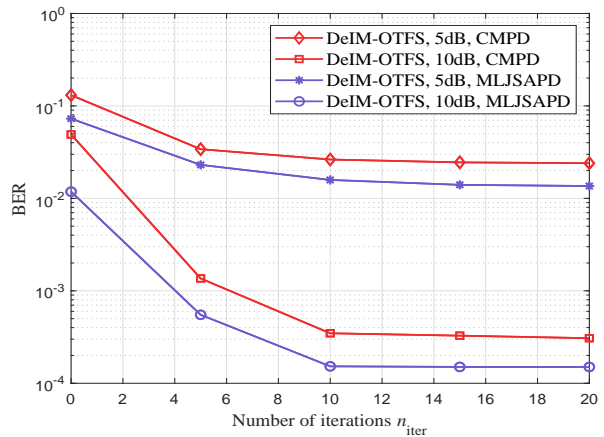}
	\caption{BER convergence comparison of the proposed MLJSAPD and CMPD algorithms at SNR = 5~dB and 10~dB with DeIM-OTFS scheme.}
	\label{fig:fig8}
\end{figure}

In Fig.~\ref{fig:fig8}, we illustrate the convergence analysis of the proposed MLJSAPD and CMPD algorithms for different SNRs with the DeIM-OTFS scheme. As shown in Fig.~\ref{fig:fig8}, the proposed MLJSAPD and CMPD algorithms at low SNR exhibit a slightly faster convergence speed than that at high SNR. At an SNR of 5~dB, the MLJSAPD and CMPD algorithms converge after 8 iterations on average. However, for high SNR of 10~dB, the MLJSAPD and CMPD algorithms converge in 10 iterations. Based on the above analysis, we take the number of iterations to be 10 for the following simulation tests. Similar convergence result can be observed for the DoIM-OTFS scheme, and thus we omit the details here for brevity. 

\begin{figure}
	\center
	\includegraphics[width=3.5in,height=2.4in]{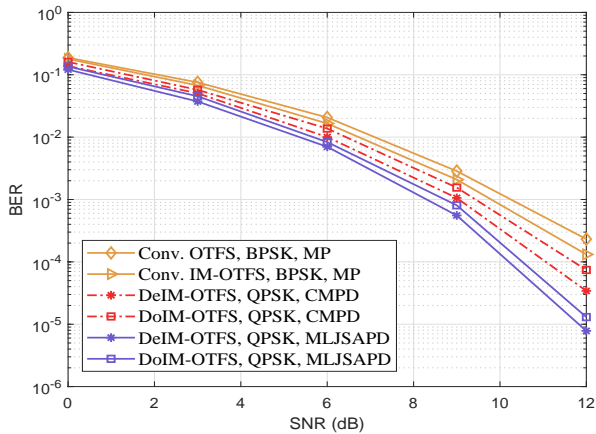}
	\caption{BER performance comparison between the proposed MLJSAPD and CMPD algorithms with DeIM-OTFS/DoIM-OTFS scheme and the traditional OTFS/IM-OTFS systems.}
	\label{fig:res_fig9}
\end{figure}

In Fig.~\ref{fig:res_fig9}, we compare the BER performance of the proposed MLJSAPD and CMPD algorithms with the DeIM-OTFS/DoIM-OTFS scheme and the traditional OTFS/IM-OTFS system. From Fig.~\ref{fig:res_fig9}, it can be observed that the BER performance of the proposed MLJSAPD algorithm for the DeIM-OTFS and DoIM-OTFS schemes are better than those of the CMPD and traditional OTFS system. Since the activation states of all resource units in each block are connected, the proposed MLJSAPD algorithm adds a new layer in the factor graph to fully exploit this prior message through each iteration, which helps explain why the output extrinsic information of the indicator nodes becomes more reliable. Therefore, the MLJSAPD algorithm can achieve better performance than the CMPD algorithm thanks to its more accurate estimation of the activated resource units. Specifically, the MLJSAPD algorithm with the DeIM-OTFS scheme shows an SNR gain of nearly 2~dB over the traditional OTFS system in the high SNR region. To demonstrate this superiority, we also compare the performance of our proposed DeIM-OTFS/DoIM-OTFS schemes to that of the existing random IM-OTFS system. It is observed that our proposed MLJSAPD and CMPD algorithms with the DeIM-OTFS/DoIM-OTFS scheme provide better performance than the conventional random IM-OTFS system. Moreover, the proposed DeIM-OTFS scheme achieves better error performance than the DoIM-OTFS scheme, which is consistent with the analysis of Fig.~\ref{fig:fig5}.

\begin{figure}
	\center
	\includegraphics[width=3.5in,height=2.4in]{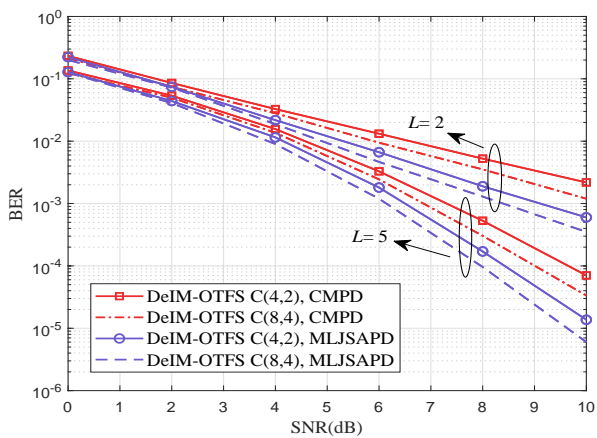}
	\caption{BER performance comparison for different activated indices of the proposed DeIM-OTFS system under different number of channel multipaths with BPSK modulation.}
	\label{fig:fig10}
\end{figure}
Fig.~\ref{fig:fig10} shows the BER performance for the DeIM-OTFS scheme with parameters ``$\widehat{M}=\widehat{N}=4,~\widehat{k}=2$" and ``$\widehat{M}=8,~\widehat{N}=4,~\widehat{k}=4$" under different numbers of channel multipaths. The SEs of ``DeIM-OTFS scheme C(4,~2)" and ``DeIM-OTFS scheme C(8,~4)" are 1.125 and 1.188bps/Hz, respectively. As seen from Fig.~\ref{fig:fig10}, although the SE of ``DeIM-OTFS scheme C(8,~4)'' is slightly higher than that of ``DeIM-OTFS scheme C(4,~2)", it can obtain better BER performance. This is due to the fact that the DeIM-OTFS scheme with C(8,~4) is more robust to the interference caused by the channel than that of C(4,~2).  
Moreover, the DeIM-OTFS scheme of $L=5$ can provide superior performance to that of $L=2$ since more diversity can be exploited from a larger number of independent resolvable paths.

\begin{figure}
	\center
	\includegraphics[width=3.5in,height=2.4in]{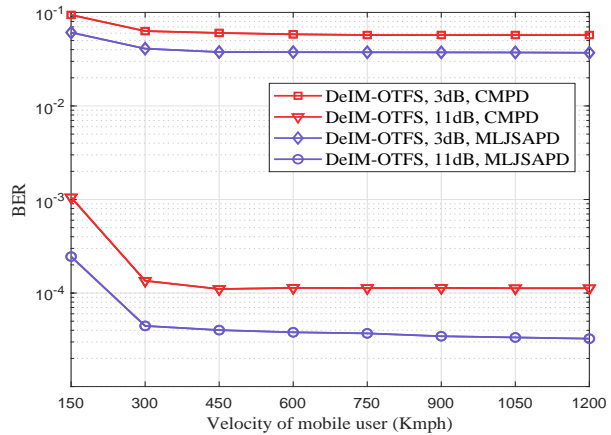}
	\caption{BER performance of MLJSAPD and CMPD algorithms for different mobile velocities.}
	\label{fig:fig11}
\end{figure}

Fig.~\ref{fig:fig11} shows the BER performance of MLJSAPD and CMPD algorithms for different user velocities with SNR = 3~dB and 11~dB. It can be observed that the BER performance of the MLJSAPD and CMPD algorithms gradually improve as the velocity increases and are saturated after the velocity beyond 450 Kmph. The underlying reason is that, as the increase of velocity, OTFS modulation can resolve higher contrast paths in the Doppler domain, and thus better performance can be achieved. As a result, it is obvious that performance improvements can be obtained at high user velocities.

\begin{figure}
	\center
	\includegraphics[width=3.5in,height=2.4in]{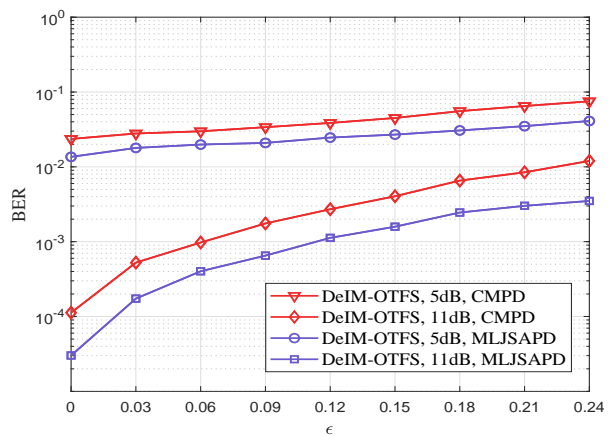}
	\caption{BER performance of MLJSAPD and CMPD algorithms with imperfect CSI.}
	\label{fig:fig12}
\end{figure}

Finally, the BER performance of the proposed MLJSAPD and CMPD algorithms are tested in terms of imperfect CSI in Fig.~\ref{fig:fig12}. Here, we characterize the CSI errors by adopting the following model
\cite{9354639}

\hspace{20mm}$h_{i}=\tilde{h}_{i}+\Delta h_{i},\left \| \Delta h_{i} \right \|\leq \epsilon_{ h_{i}}$,

\hspace{20mm}$\nu _{i}=\tilde{\nu}_{i}+\Delta \nu _{i},\left \| \Delta \nu _{i} \right \|\leq \epsilon_{ \nu _{i}}$,

\hspace{20mm}$\tau_{i}=\tilde{\tau}_{i}+\Delta \tau_{i},\left \| \Delta \tau_{i} \right \|\leq \epsilon_{ \tau_{i}}$,

\hspace{-2.5mm}where $\tilde{h}_{i}$, $\tilde{\nu}_{i}$ and $\tilde{\tau}_{i}$ denote the estimated values of $h_{i}$, $\nu _{i}$, and $\tau_{i}$, respectively. $\Delta h_{i}$, $\Delta \nu _{i}$, and $\Delta \tau_{i}$ are the corresponding channel estimation errors. We assume the norms of $\Delta h_{i}$, $\Delta \nu _{i}$, and $\Delta \tau_{i}$ do not exceed the given values of $\epsilon_{ h_{i}}$, $\epsilon_{ \nu _{i}}$ and $\epsilon_{ \tau_{i}}$, respectively. Here, we set $\epsilon_{ h_{i}}=\epsilon\left \| \tilde{h}_{i} \right \|$, $\epsilon_{ \nu _{i}}=\epsilon\left \| \tilde{\nu}_{i} \right \|$ and $\epsilon_{ \tau_{i}}=\epsilon\left \| \tilde{\tau}_{i} \right \|$ for simplicity. From Fig.~\ref{fig:fig12}, we can observe that both the proposed MLJSAPD and CMPD algorithms only suffer from a mild performance loss for the modest values of channel uncertainty $\epsilon$. With the increase of the level of channel uncertainty, the rapid degradation in BER performance does not appear, which verifies the robustness of our proposed MLJSAPD and CMPD detection algorithms.


\section{Conclusion}
In this paper, we have proposed two efficient block-wise IM schemes for practical high mobility OTFS communications, namely DeIM-OTFS and DoIM-OTFS. We have analyzed the average BER bounds for the proposed DeIM-OTFS and DoIM-OTFS schemes with the optimal ML detectors. Both theoretical analysis and simulation results have demonstrated that our proposed DeIM-OTFS and DoIM-OTFS schemes outperform the conventional random IM-OTFS scheme. We have also noted that our proposed DeIM-OTFS scheme outperform the DoIM-OTFS scheme as the interference effect caused by the channel delays is much less than that caused by the channel Doppler spreads.
Furthermore, we have developed low-complexity MLJSAPD and CMPD algorithms for symbol detection in the proposed DeIM-OTFS and DoIM-OTFS systems. Numerical results have verified that our proposed MLJSAPD and CMPD algorithms can achieve desired performance and robustness to the imperfect CSI. The proposed MLJSAPD algorithm can achieve superior performance to the CMPD algorithm with a slight sacrifice in complexity.

\bibliographystyle{IEEEtran}
\bibliography{reference}

\end{document}